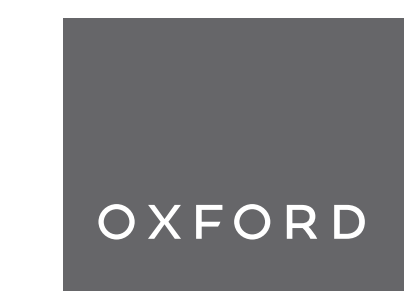

# Human–AI Interactions in Public Sector Decision Making: "Automation Bias" and "Selective Adherence" to Algorithmic Advice

**Saar Alon-Barkat,*** **Madalina Busuioc†**

*University of Haifa, Israel
†Vrije Universiteit Amsterdam, The Netherlands
Address correspondence to the author at e.m.busuioc@vu.nl.

## Abstract

Artificial intelligence algorithms are increasingly adopted as decisional aides by public bodies, with the promise of overcoming biases of human decision-makers. At the same time, they may introduce new biases in the human–algorithm interaction. Drawing on psychology and public administration literatures, we investigate two key biases: overreliance on algorithmic advice even in the face of "warning signals" from other sources (*automation bias*), and selective adoption of algorithmic advice when this corresponds to stereotypes (*selective adherence*). We assess these via three experimental studies conducted in the Netherlands: In study 1 (*N* = 605), we test *automation bias* by exploring participants' adherence to an algorithmic prediction compared to an equivalent human-expert prediction. We do not find evidence for automation bias. In study 2 (*N* = 904), we replicate these findings, and also test *selective adherence*. We find a stronger propensity for adherence when the advice is aligned with group stereotypes, with no significant differences between algorithmic and human-expert advice. In study 3 (*N* = 1,345), we replicate our design with a sample of civil servants. This study was conducted shortly after a major scandal involving public authorities' reliance on an algorithm with discriminatory outcomes (the "childcare benefits scandal"). The scandal is itself illustrative of our theory and patterns diagnosed empirically in our experiment, yet in our study 3, while supporting our prior findings as to automation bias, we do not find patterns of selective adherence. We suggest this is driven by bureaucrats' enhanced awareness of discrimination and algorithmic biases in the aftermath of the scandal. We discuss the implications of our findings for public sector decision making in the age of automation. Overall, our study speaks to potential negative effects of automation of the administrative state for already vulnerable and disadvantaged citizens.

## Introduction

Artificial intelligence (AI) algorithms are being widely adopted in the public sector across jurisdictions. Essentially a set of tools that display (or can even surpass) human-level performance on given tasks traditionally associated with human intelligence, AI algorithms are being relied upon in areas as varied as policing, welfare, criminal justice, healthcare, immigration, or education (Busuioc 2021; Calo and Citron 2021; Diakopoulos 2014; Eubanks 2018; Engstrom et al. 2020; O'Neil 2016; Richardson, Schultz, and Crawford 2019; Veale and Brass 2019; Yeung and Lodge 2019), increasingly permeating non-routine and high-stakes aspects of bureaucratic work. The growing and deepening reliance on AI and machine learning technologies in the public sector has been diagnosed as "transformative" of public administrations (Bullock 2019; Vogl et al. 2020; Young, Bullock, and Lecy 2019).

These developments are driven by the promise of policy solutions that are potentially more effective, efficient, and low-cost. In addition, and importantly, algorithms are said to come with the "promise of neutrality," in contrast to decision making based on human intuition, which involves biases and can result in discrimination. In other words, AI use in decision making is said to hold the potential to help us overcome our cognitive biases and limitations. This has been an important driver for the adoption of such technologies in highly consequential public sector areas such as law enforcement or criminal justice: Predictive policing technologies, for instance, were propagated in the US context "as one answer to racially discriminatory policing, offering a seemingly race-neutral, 'objective' justification for police targeting of poor communities" (Ferguson 2017, 5). Numerous other jurisdictions have followed suit with predictive technologies relied upon by police forces in the United Kingdom, the Netherlands, Germany, among many others. Like rationales precipitated the adoption of predictive risk assessment systems in criminal justice, similarly in part in response to concerns with human bias and discrimination (Israni 2017), despite such systems themselves being flagged as sources of bias (Angwin et al. 2016).

For a large part, AI algorithms currently serve as *decisional aides* to human decision-makers ("decision-support systems") in many bureaucratic contexts. This is especially so in highly consequential public sector areas, where "full automation seems inappropriate or far off" (Edward and Veale 2017, 45). Rather than making decisions on their own, algorithmic outputs—be they risk assessment scores used in criminal justice or the algorithm-generated "heat maps" of

We thank the three anonymous JPART reviewers for their extremely helpful comments. We are also grateful to Dimiter Toshkov, Nadine Raaphorst, Joris van der Voet, Dana Vashdi, Sharon Gilad, Markus Tepe, Stephan Grimmelikhuijsen, Thijs de Boer, Benjamin Tidå, Omer Yair, and Aaron Swaving for their valuable comments and feedback in the process of developing this project. We thank Luuk van Roozendaal for excellent research assistance.

predictive policing—support human decision making. As such, algorithms do not remove the human decision-maker out of the equation—instead, algorithmic decision making arises at the interaction of the two.

For all its promise, the deployment of AI algorithmic technologies in the public sector has raised important concerns. High among these are concerns with algorithmic accountability and oversight of algorithmic outputs (Busuioc 2021; Diakopoulos 2014); issues of "algorithmic bias"—the well-documented propensity of algorithms to learn systemic bias through, among others, their reliance on historical data and come to perpetuate it, effectively "automating inequality" (Eubanks 2018); as well as the potential for bias arising from human *processing* of AI algorithmic outputs. This article focuses on the latter, which we believe is an important and especially worthy aspect of analysis in light of algorithms' roles as decisional aids in public sector decision making. In this context, it becomes important to understand the implications of these technologies in shaping public sector decision making and specific cognitive biases that might arise in this respect. This gains yet further relevance as in the context of the rise of algorithmic governance, human decision-makers are regarded as important safeguards, as decisional mediators, on issues of algorithmic bias. Investigating to what extent our cognitive limits allow us to act as effective decisional mediators becomes critical in an increasingly automated administrative state.

In this article, we focus on *two diverging biases*, theorizing on the basis of two strands of literature from different disciplines that have thus far not spoken to each other on this topic. The first bias, which builds on previous social psychology studies is *automation bias*. It refers to a well-documented human propensity to automatically defer to automated systems, despite warning signals or contradictory information from other sources. In other words, human actors are found to uncritically abdicate their decision making to automation. While robust, these findings have been documented for AI algorithmic precursors such as pilot navigation systems and in fields outside a public sector context. The second bias we theorize and test can be extrapolated from existing public administration research on biased information processing, and pertains to decision-makers' *selective adherence* to algorithmic advice. Namely, the propensity to adopt algorithmic advice selectively, when it matches pre-existing stereotypes about decision subjects (e.g., when predicting high risk for members of negatively stereotyped minority groups). This bias has not yet been investigated in our field with respect to algorithmic sources.

We report the results of three survey experiment studies conducted in the Netherlands, which provide rigorous tests for these hypothesized biases. In study 1 ($N$ = 605), we put automation bias to test by exploring participants' adherence to an algorithmic prediction (which contradicts additional evidence) and comparing it to an equivalent human-expert prediction. In study 2 ($N$ = 904), we replicate these findings, and at the same time, we also test whether decision subjects' ethnic background moderates decision-makers' inclination to follow the algorithmic advice. In other words, whether respondents are more likely to follow an algorithmic advice when this prediction is aligned with pre-existing group stereotypes (engaging in "selective" rather than automatic adherence). Studies 1 and 2 were conducted among citizens in a context where citizens can act as decision-makers. In study 3, we set out to further replicate our findings with a sample of Dutch civil servants ($N$ = 1,345). During our preparations for that study, a major political scandal occurred in the Netherlands (the "childcare benefits scandal"), involving algorithm use by public authorities. The scandal involved tax authorities' reliance as a decisional aid on an AI algorithm that used nationality as a discriminant predictive feature, with ensuing bureaucratic decisions reflecting discrimination of minority groups. We discuss the results of study 3 in light of its co-occurrence with these events, which closely align with our theory.

Our focus is on human processing biases arising from the use of AI algorithms in a *public sector context*. While we would expect such biases to be equally relevant for algorithmic decision making in the private sector, we focus on the public sector because the stakes are especially high for governments. AI algorithms are increasingly adopted in high-stakes areas—where they are highly consequential for individual's lives, rendering these questions especially pressing in a public sector context.

## Automation and Decision Making in the Public Sector: A Tale of Two Biases

An important and growing literature in public administration is concerned with the effects of the increasing reliance on digital technologies for public sector decision making. A key concern in particular pertains to the implications of these technologies for the discretion and professional judgment of decision-makers such as (street-level) bureaucrats (Bovens and Zouridis 2002). This literature has flagged the potential, in the age of automation, for "digital discretion" (Busch and Henriksen 2018), "automated discretion" (Zouridis, van Eck, and Bovens 2020) or specifically in the context of AI, "artificial discretion" (Young, Bullock, and Lecy 2019) to supplant the discretion of bureaucratic actors in the administration (see also Buffat 2015; Bullock 2019; de Boer and Raaphorst 2021). In other words, the potential of digital tools to "influence or replace human judgment" in public service provision (Busch and Hendriksen 2018, 4) and to alter the very nature of public managers' work (Kim, Andersen, and Lee 2021) and bureaucratic structures and routines (Meijer, Lorenz, and Wessels 2021). Such tools stand to fundamentally shape public sector decision making through constraining, or even removing, the scope for human expertise and discretion or influencing human judgment and cognition in unexpected ways. In doing so, the delegation of administrative decision-making authority to AI technologies could have profound implications for bureaucratic legitimacy (Busuioc 2021) and public values more broadly (Schiff, Schiff, and Pierson 2021).

In this context, it becomes important therefore to understand how decision-makers in a public sector context process algorithmic outputs used as decisional aids, how they incorporate them into their decision making, their implications and whether these differ in significant ways from the processing of traditional (human-sourced) advice. To operationalize the potential implications of AI advice for decision making, and given limited theorizing of potential cognitive biases in this emerging area, we borrow from, theorize and integrate insights from two separate strands of literature, which offer important starting points to unpack this topic: social psychology literature on automation and public administration research on information processing. Interestingly, these two literatures offer us somewhat competing projections as to what to expect.

## Automation Bias: Automatic Adherence to Algorithmic Advice

While AI is meant to help us overcome our biases, research from social psychology suggests that automated systems might give rise to new and distinct biases arising from human processing of automated outputs. "Automation bias" is a well-recognized decisional support problem that has emerged from studies in aviation and healthcare, areas that have traditionally heavily relied on automated tools. Automation bias refers to undue deference to automated systems by human actors that disregard contradictory information from other sources or do not (thoroughly) search for additional information (Cummings 2006; Lyell and Coiera 2017; Mosier et al. 2001; Parasuraman and Riley 1997; Skitka, Mosier, and Burdick 1999, 2000; Skitka et al. 2000). In other words, it is manifest in the "the use of automation as a heuristic replacement for vigilant information seeking and processing" (Mosier et al. 1998, 201), a "short cut that prematurely shuts down situation assessment" (Skitka, Mosier, and Burdick 2000, 714).

Experimental lab studies have diagnosed this tendency across a number of research fields (Goddard, Roudsari, and Wyatt 2012). While robust, these findings have not been investigated *in a bureaucratic context*. As such, we do not know to what extent such biases are relevant and replicate in administrative contexts. Extant studies suggest that this propensity to defer to automation stems on the one hand, from the perceived inherent superiority of automated systems by human actors and on the other, from "cognitive laziness," a human reluctance to engage in cognitively demanding mental processes, including thorough information search and processing (Skitka, Mosier, and Burdick 2000, 702). Research findings on automation bias are further supported by ample anecdotal evidence of automation bias with respect to commercial flights (Skitka et al. 2000, 703), car navigation systems (Milner 2016) and more recently, also specifically documented in the context of AI for self-driving cars (National Transportation Safety Board 2017). Recent business management experiment-based studies similarly talk about "algorithm appreciation" (Logg, Minson, and Moore 2019), describing a similar tendency to over-trust algorithmic outputs.

Concerns with *automation bias* have been increasingly voiced by scholars in the context of a growing reliance on AI tools in the public sector and high-stakes scenarios (Cobbe 2019; Edwards and Veale 2017; Medium—Open Letter Concerned AI Researchers 2019; Zerilli et al. 2019), and increasingly so also by public administration scholars (Busuioc 2021; Peeters 2020; Giest and Grimmelikhuijsen 2020; Young et al. 2021). More broadly, this also corresponds to concerns raised by public administration literature, as discussed above, on the potential of AI algorithmic tools (and digital tools more broadly) to replace bureaucratic discretion and professional judgment. Our investigation into automation bias speaks directly to this literature through setting out to elucidate whether the scope for discretion of human decision-makers is removed through the introduction of such tools.

Such concerns become particularly relevant given well-documented failures and malfunctioning of AI(-informed) systems (e.g., Benjamin 2019; Buolamwini and Gebru 2018; Ferguson 2017; Eubanks 2018; O'Neill 2016; Richardson, Schultz, and Crawford 2019; Rudin 2019). Due, among others, to model and/or data inadequacies, AI algorithms have been found to reproduce and automate systemic bias, and to do so in ways that, by virtue of their opaqueness and/or high complexity, have proven difficult to diagnose for both domain experts and system engineers alike. A human propensity for default deference to algorithmic systems under such circumstances would become especially problematic—even more so given the high-stakes of AI use in a public sector context.

> $H_1$—Decision-makers are more likely to trust and to follow algorithmic advice than human advice, when faced with similar contradicting external evidence. (automation bias)

## Selective Adherence to Algorithmic Advice

We theorize a second, diverging concern regarding decision-makers' use of algorithmic advice extrapolating from behavioral work on public decision-makers' *information processing*. Following a *motivated reasoning* logic, this growing body of literature has established that decision-makers are prone to *selectively* seek and interpret information in light of pre-existing stereotypes, beliefs, and social identities. They assign greater weight to information congruent with prior beliefs and contest inputs that contradict them (Baekgaard et al. 2019; Baekgaard and Serritzlew 2016; Christensen et al. 2018; Christensen 2018; James and Van Ryzin 2017; Jilke 2017; Jilke and Baekgaard 2020). These studies have demonstrated these "confirmation biases" with regards to the processing and interpretation of "unambiguous" information such as performance indicators. However, this has *not* been explicitly theorized nor investigated yet in relation to algorithmic decisional aides.

We theoretically extend this literature, and argue that this motivated reasoning mechanism would apply not only to information inputs generated by humans, but also to information produced by AI algorithms. Thus, we would similarly expect decision-makers to adhere to algorithmic advice *selectively*, when it matches stereotypical views of the decision subject (rather than by default, as expected by automation bias literature). This theoretical expectation also corresponds to works on bureaucratic discrimination indicating that bureaucratic decision-makers search for stereotype-consistent cues in their decisions, or respond to them unconsciously (e.g., Andersen and Guul 2019; Assouline, Gilad, and Bloom 2022; Jilke and Tummers 2018; Pedersen et al. 2018; Schram et al. 2009). In this regard, we theorize that an algorithmic prediction that accords with a group stereotype would similarly amount to such a cue, which provides expectancy confirmation.

While public administration scholars have thus far not investigated selective processing of algorithmic outputs, it has been the subject of recent investigations by law and computer science scholars in studies on the use of algorithmic risk assessment by criminal courts (Green and Chen 2019a, 2019b; Stevenson 2018), diagnosing patterns that are consistent with selective adherence and motivated reasoning.

Hence, extrapolating from and theorizing on the basis of these literatures we first hypothesize that:

> $H_2$—Decision-makers are more likely to follow advice (human or algorithmic-based) that matches stereotypical views of the decision subjects. (selective adherence)

To clarify, $H_2$ pertains to the expectation that selective adherence biases diagnosed for human-sourced advice are also present for algorithmic advice, that is, *selective adherence, across both human and algorithmic advice types*. In other words, we theorize that these biases persist (do not disappear) in the adoption of algorithms in public sector decision making. Establishing whether selective adherence is present is important in a context where AI algorithms are said to have the potential to do away with human decisional biases. What is more, the presence of selective adherence biases gains special relevance in the algorithmic case. As evidence of systematic algorithmic biases is accumulating, human decision-makers in-the-loop are seen as critical checks, in their roles as decisional mediators. Investigating the presence of selective adherence, importantly, therefore, also speaks to the extent to which human decision-makers can actually function as effective decisional mediators and safeguards against such risks.

If selective adherence biases are to persist, the next question is whether they are more emphasized in the use of algorithms. Are decision-makers *more prone* to *selective adherence* to algorithmic advice compared to equivalent human advice? In other words, do algorithmic outputs *exacerbate* the risk of selective adoption and discriminatory decisions? We theorize that algorithms have the potential to amplify these biases due to their unique nature. Literature on automation has theorized that automated decisional aids tend to create a "moral buffer," acting as a psychological distancing mechanism resulting in a diminished sense of moral agency, personal responsibility and accountability for the human actor "because of a perception that the automation is in charge" (Cummings 2006, 8). These feelings of moral and ethical disengagement and decreased responsibility may reduce decision-makers' awareness of potential biases and implicit prejudice. Or worse: the algorithmic advice could vindicate and give free license to decision-makers' latent views (racial, xenophobic, misogynistic, etc.) by providing them with a seemingly legitimate reason to adopt discriminatory decisions. Algorithms, in other words, could serve to "give permission" to decision-makers to act on their biases: Algorithms' face-value "neutral" or "objective" character would fend-off potential suspicions of bias and/or confirm the validity of biased or prejudiced decisions. An algorithmic recommendation aligned with decision-makers' own biases could amount to a powerful (mathematical!) endorsement thereof. We therefore expect biased adherence to become especially emphasized for algorithmic advice by comparison with human advice.

Consequently, we further hypothesize that:

> $H_3$—Selective adherence is likely to be exacerbated when decision-makers receive an algorithmic rather than a human advice. (exacerbated selective adherence)

## Empirical Evidence from Previous Studies

To date, we lack systematic empirical evidence about the prevalence of cognitive biases in algorithm-based public sector decision-making. Existing peer-reviewed empirical studies on this topic are from law and computer science scholars in the context of algorithm use in pretrial criminal justice decisions in the United States. These studies stem from the underlying concern with high levels of detention in the United States and its growing carceral state and are aimed at investigating the promise of algorithmic risk assessments to decrease detention levels through improving the accuracy of judges' assessments of recidivism risk. Their tentative findings, as detailed below, are consistent with our theorized patterns of *selective adherence*. Stevenson (2018) uses archival data of criminal cases from the state of Kentucky to compare observationally detention rates before and after a reform in 2011 that made risk assessment mandatory in pretrial procedures. She finds that the expansion in the use of risk scores led to an overall increase in pretrial release immediately after the implementation of the reform, however, this eroded and almost disappeared within a matter of years. Additionally, the study finds that judges were more likely to accept low scores for white defendants, while overriding similar scores for black defendants.

These findings are further supported by a series of experimental studies among laypersons (Green and Chen 2019a, 2019b; Grgić-Hlača, Engel, and Gummadi 2019). These studies include a judicial decision-making task in which participants are shown details of arrests and are asked to predict recidivism risk, comparing participants' predictions with/without an algorithmic risk assessment. Grgić-Hlača, Engel, and Gummadi (2019) find that participants did not significantly change their decisions in response to the algorithmic prediction, even when they receive feedback about its high accuracy or are incentivized to make correct predictions. Green and Chen (2019a, 2019b) further compare between outcomes for black and white defendants and diagnose participant reliance on algorithms indicative of "disparate interactions": participants adhered to the algorithmic advice to a greater degree when it predicted either high risk for a black defendant or low risk for a white defendant.

All in all, while most of these studies demonstrate that public decision-makers can be affected in their decisions by algorithmic decisional aids, they do not provide particularly strong evidence for automatic deference to algorithmic advice, as would be expected on the basis of automation bias literature. They provide instead tentative empirical evidence that decision-makers tend to process such advice in a biased, selective manner.

Still, these studies have several important limitations. *First*, while the aim of these studies was to learn about the influence of algorithmic decisional aids, their comparison was only to a condition where decision-makers did not receive any advice at all, as opposed to comparable human expert advice. It is an open question, therefore, whether the effects found are attributed to algorithms *per se*, or rather that other professional advice that similarly includes numeric outputs would yield the same outcome. We propose that in order to isolate the distinct effect of algorithms, the appropriate counterfactual should be an equivalent numeric advice produced by a human expert. *Second*, we argue that these studies are ill-equipped to investigate automation bias, since they lacked additional contradictory evidence or inputs from other sources. Rather, automation bias can be tested effectively by supplementing the algorithmic advice with such additional inputs, a condition which "forces" decision-makers to choose whether to rely on the automated authority or rather take into account additional information and indicators. A similar approach was applied by previous automation bias experimental studies, where participants were given automated aids not aligned with other indicators (Mosier et al. 1998; Skitka, Mosier, and Burdick 1999, 2000; Skitka et al. 2000). *Thirdly*, these studies are focused on the application of algorithms in one specific policy context. It is important to explore the generalizability of these patterns to additional public policy areas, especially given the rapid spread of algorithms across various policy contexts and jurisdictions.

Below, in the methodology section, we present our unique research design, and discuss how it overcomes these limitations.

## Research Design

To examine our hypotheses, we designed and conducted a series of three unique survey experiments among Dutch citizens and civil servants. Study 1 ($N$ = 605) was designed to test our *automation bias* hypothesis. Study 2 ($N$ = 904) was designed to replicate study 1 on a separate sample, as well as to test our two hypotheses regarding *selective adherence* to algorithmic advice. Studies 1 and 2 were conducted among Dutch citizens in a context where citizens can act as decision-makers. Thereafter, in study 3 ($N$ = 1,345), we repeated our experimental design with a large sample of Dutch civil servants. The demographic characteristics of our samples are summarized in Appendix 2.

The studies involve an administrative decision-making task that concerns local school board decisions on the employment of teachers. As elaborated below, we utilized a hypothetical scenario of an algorithmic performance evaluation tool, used as a decisional aid for the assessment of Dutch high-school teachers.

In the Netherlands, members of such boards are not required to complete a specific professional certification, and are composed, among others, of volunteers (lay persons) such as parents or citizens from the local community (OECD 2014a, 14; OECD 2014b, 22, 98). As such, lay citizens are relevant decision-makers in this context. Moreover, to further enhance the external validity of our study, we additionally replicate the study with a large sample of actual civil servants—Dutch decision-makers from various policy areas and across government levels. An important advantage of our choice of empirical setting is that it involves a bureaucratic task that can be relatively easily exercised in a vignette survey experiment with participants who are not necessarily experts on the specific task, allowing us to test our expectations among decision-makers in a public sector context more broadly. Our explicit aim is to tap into generalized human biases in algorithm-supported decision making in the public sector.

Our decision to focus on the education setting in the vignette was inspired by the real-life case of Sarah Wysocki—a teacher in the United States who was fired based on the prediction of an algorithmic score, while ignoring her record and reputation as a well-performing teacher (Turque 2012). Wysocki's story is often mentioned as an illustrative example as to the dangers of bureaucracies' reliance on black-box algorithms (see O'Neal 2016). We aimed to simulate a similar scenario in which officials are required to make a decision of whether to extend the employment contract of a teacher, when an algorithmic score indicates that she performs poorly, yet additional evidence suggests otherwise. We test experimentally whether people are more inclined to adhere to such advice when produced by an algorithm, compared to a human expert, as expected by our automation bias hypothesis. We further examine (in studies 2 and 3) whether participants are more likely to follow such advice when it concerns a decision subject from an ethnic minority background, and whether participants do so to a greater extent when the advice comes from an algorithm (as opposed to a human expert). This allows us to explore instead patterns of selective (rather than automatic) adherence.

We tailored our survey experimental design to the Dutch context. In the Netherlands, all schools operate under publicly funded educational associations, which enjoy a large autonomy in their management. Important decisions, including personnel management, are made by a school board, which includes representatives of the educational association. In our study, as detailed below, we invite participants to a simulation task where they act as board members of a hypothetical Dutch high-school and are asked to make decisions about the employment of three new teachers. Below we present each of the three studies and their results. In addition, the results are summarized in supplementary table A4.

## Study 1: Automatic Adherence to Algorithmic Versus Human Advice (Automation Bias)

Study 1 is designed to examine our hypothesis that decision-makers are inclined to over-trust algorithmic advice—to follow algorithmic predictions despite additional contradicting evidence, and to do so to a greater extent than when presented with equivalent advice by a human expert ($H_1$). We preregistered the study and administered it in February 2020.[1] The survey experiments were hosted on Qualtrics, and participants ($N$ = 605) were recruited through a large online panel company—Dynata.[2] The survey was conducted in Dutch.

### Procedure

Survey participants are asked to act as board members of a hypothetical Dutch high school. In the main experimental task, we ask participants to make a decision regarding the employment of three teachers, who were hired the previous year for a trial period. Only two of the three new teachers can be permanently hired and accordingly, participants must choose *one* teacher whose contract will *not* be renewed. As a basis for their decision, participants are given *two data inputs* per teacher (one qualitative input, and one numeric input—a score) in both the algorithmic and the human advice conditions. In the *algorithmic condition*, respondents are told the numeric input is produced by an algorithm, while for the *human-expert advice condition* that it is produced by a human expert.

[1]The preregistration form of study 1 is available at https://aspredicted.org/5de9d.pdf. Methodological choices are further discussed in the supplementary section A6.

[2]We estimated that a modest effect size of OR = 1.5 is detectable with power of 0.8 ($p$ = .05, one-sided test), assuming a probability of .3 for the baseline human-advice group.

The *first* input, which was identical for all participants, is a brief summary of a qualitative evaluation by the HR person of the educational association. The *second* is a numeric prediction of the teachers' potential to perform well in the future, ranging between 1 (lowest) to 10 (highest). Participants are told that this numeric prediction was conducted by a body named ILE (short for "Innovatieve Lerarenevaluatie"—"Innovative Teachers' Evaluation"), and accordingly we refer to it as the "ILE evaluation score". Respondents are randomly assigned to one of two conditions: they are told that the ILE score is either produced by a machine learning algorithm (*algorithmic advice condition*), or by consultants (*human-expert advice condition*). To bolster participants' confidence in the predictive capacity of the ILE score, we noted (in both conditions) that it "has proven highly effective in predicting teacher performance, with an accuracy rate of 95%."

It is noteworthy that the format we used for the ILE evaluation score (an integer between 1 and 10) was designed to resemble the COMPAS risk score that is used in pretrial procedures across the United States, which similarly ranges from 1 to 10. The comparison between a numeric algorithmic prediction and additional qualitative evidence (e.g., case file evidence presented to a judge) is typical for many policy areas where algorithms are used as decisional aids.

Participants were shown a table that presents the three teachers and the two inputs for each teacher, as illustrated in figure 1. To minimize additional differences in the characteristics of the three teachers, which could potentially affect participants' decisions, all three teachers are female, have typical Dutch names and their teaching areas are in natural sciences. The order of the three teachers was randomized (see also supplementary section A5.4).

In line with our theoretical focus, we deliberately designed the task so that there will be an *incongruence* between the two inputs in the table: the lowest ILE score (4) is never matched with the most negative qualitative HR evaluation. The incongruence was as follows: One of the three teachers received a low ILE score of 4, whereas the other two received scores of 8 and 6. The HR person's qualitative evaluation similarly varies as one of the three teachers gets negative remarks, whereas the other two teachers receive positive and respectively, mixed evaluations. Most importantly however, the negative qualitative evaluation is never assigned to the teacher with the lowest ILE score (4), but to one of the other teachers. Instead, the teacher with the lowest ILE score receives either the positive or the mixed qualitative evaluation. Accordingly, participants faced a decision of whether or not to follow the ILE score (i.e., to fire the teacher with the most negative ILE score), given its incongruence with the HR person's qualitative evaluation.

For exploratory purposes, we also randomized the distribution of the ILE scores (4, 6, and 8) across the three teachers, to generate different levels of incongruence between the ILE score and the qualitative evaluation. We assigned participants to one of two main conditions of incongruence. In the high incongruence condition (displayed in figure 1), the teacher with the lowest ILE score receives the most favorable qualitative evaluation. In the modest incongruence condition, the teacher with the lowest ILE score receives the mixed qualitative evaluation. In other words, through the qualitative input (the HR evaluation), respondents in both conditions receive informational cues that are at odds, to varying degrees, with the ILE score.

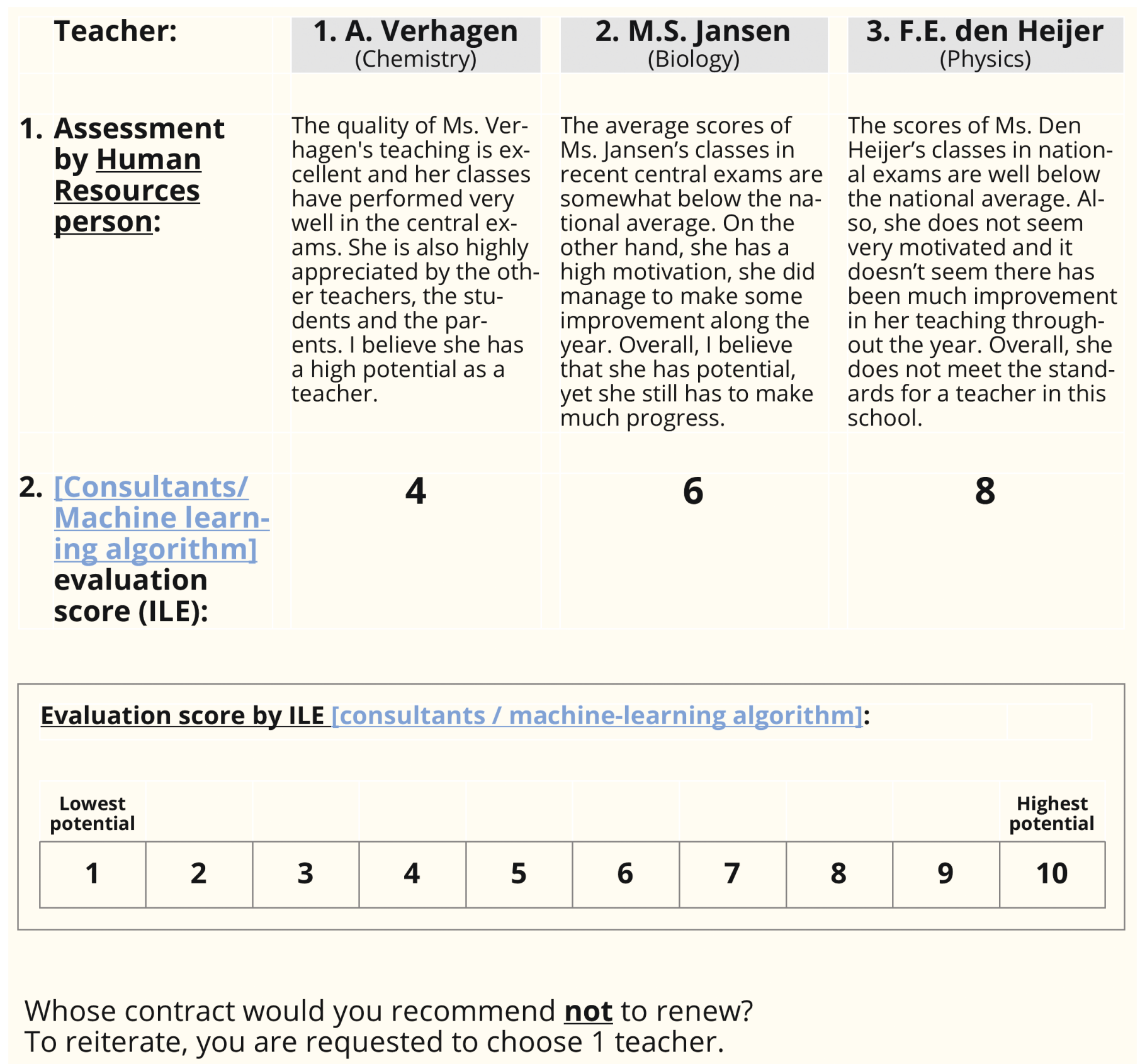

**Figure 1.** Illustration of the Experimental Task

Our main *outcome variable* is participants' likelihood to follow the ILE score. We coded 1 when participants chose to fire the teacher with the lowest ILE score and 0 otherwise. In our analyses below, we compare this binary variable between the two conditions (algorithmic versus human-expert).

The main task was followed by a series of manipulation check questions to confirm that participants were aware of the source of advice (algorithmic versus human) as well as of the actual ILE score (supplementary section A3). The survey further included an attention check, additional items regarding participants' perceptions of algorithms and their familiarity with the use of algorithms by public bodies and a set of demographic questions. The full survey is attached in the supplementary sections A8 and A9.

We excluded from all analyses participants who did not pass the attention check or completed the questionnaire under 3 minutes. These filtering criteria are not associated with the assignment to the experimental conditions (supplementary table A2.1). The two experimental groups are balanced in relation to gender, reported income and education, yet participants assigned to the algorithmic group are slightly older on average (supplementary table A1.1). In robust analyses, we further control for these covariates (supplementary table A5.1.1). While this sample consists of Dutch citizens, their average age (47) and the share of participants with high education (50%) are comparable to that of the population of Dutch civil servants. Compared with civil servants, our sample over-represents women, and people aged less than 25 or above 65 (Appendix 2). In our analyses below, we control for these variables and confirm that these characteristics do not affect our results.

A technical clarification on our statistical reporting: in all results tables presented in the article we use two-tailed *p* values uniformly, for consistency. We additionally report, for our preregistered *directional* hypotheses, the one-sided *p* values, both in the tables and in the main text when discussing the effects.

## Results (Study 1)

Tables 1 and 2 present the main experimental results of study 1. Table 1 reports the results of the logistic regression analysis as to the effect of our manipulation of the type of advice (algorithmic versus human) and Table 2 presents descriptively the distribution of participants' decisions across the two conditions. Based on our first hypothesis, and in line with automation bias literature, we expected the probability of following the advice of the ILE score (i.e., selecting *not* to renew the contract of the teacher with the lowest ILE score) to be higher among those assigned to the AI algorithmic advice, compared to those receiving an equivalent prediction produced by human experts.

In contrast to our theoretical expectation, we find very small, statistically insignificant differences between the algorithmic-advice and human-advice conditions (table 1). Under both conditions, the vast majority of participants chose to override the ILE score and instead preferred to fire (not renew) the teacher with the poorest qualitative evaluation (table 2).

**Table 1.** Study 1—Regression Results of Participants' Adherence to Algorithmic Versus Human-Expert Advice (Automation Bias)

| Predictors | (1) | | |
|---|---|---|---|
| | *OR* [95% CI] | *z* | *p* value |
| Algorithm | 0.96 [0.58–1.58] | −0.16 | .876 |
| Intercept | 0.14 [0.09–0.19] | −11.41 | <.001 |
| Observations | 605 | | |
| Log-likelihood | −218.769 | | |

*Note*: Logistic regression model; *OR*, odds ratio; *p* values refer to a two-sided test (by default). Binary outcome: following the ILE score (1 = non-renewal of employment of teacher with lowest ILE score).

**Table 2.** Study 1—Descriptive Results (Automation Bias)

| Outcome: Teacher Selected (Non-renewal of employment) | Algorithmic Advice (*n* = 295), % | Human-Expert Advice (*n* = 310), % |
|---|---|---|
| **Teacher with lowest ILE score (algorithmic/human-expert)** | **11.5** | **11.9** |
| Teacher with poorest qualitative evaluation | 77.3 | 81.0 |
| Other | 11.2 | 7.1 |

Including covariates and restricting the samples to those who successfully completed the manipulation checks does not change the results (supplementary tables A5.1.1 and A5.2.1).

Furthermore, these patterns are similar regardless of whether the lowest ILE score was assigned to the teacher with the best qualitative evaluation (high incongruence condition) or the teacher with the mixed qualitative evaluation (modest incongruence condition)—providing further confidence that the diagnosed patterns are stable (supplementary section A5.3). Also, randomizing the order of the three teachers did not significantly alter the results (supplementary section A5.4).

In summary, in study 1, we did not find evidence supporting the automation bias expectation. The majority of participants, under both algorithmic and human advice conditions, and across the conditions of incongruence, chose to override the ILE score.

## Study 2: Selective Adherence to Algorithmic Versus Human Advice Matching Stereotypes

The purpose of study 2 is two-fold: First, it aims to replicate the results of study 1 on a separate sample. Second, it is also designed to test the additional hypotheses that, similar to human advice, decision-makers are more inclined to follow algorithmic advice inasmuch as this is aligned with stereotypical views of the decision subjects ($H_2$), and that this selective adherence pattern is exacerbated by AI algorithms compared with equivalent human expert advice ($H_3$). We preregistered the study and administered it mid-March 2020, and similarly recruited participants through Dynata (*N* = 904).[3]

[3]The preregistration forms of study 2 is available at https://aspredicted.org/v3u29.pdf. Methodological choices are further discussed in the supplementary section A6.

## Procedure

We repeated the procedure of study 1, while adding a manipulation of teachers' names as a cue for their ethnic background. The control condition is identical to study 1—all three teachers are given typical Dutch surnames ("Verhagen," "Jansen," and "den Heijer"). In the treatment condition, the name of the teacher who received the lowest ILE score (4) is changed to "El Amrani," a common surname for citizens with a Moroccan background. We henceforth refer to these conditions as "Dutch teacher" and "Moroccan-Dutch teacher." We specifically selected this ethnic minority group in the Netherlands, since it is a minority group that is often negatively stereotyped (Jilke, Van Dooren, and Rys 2018; Kamans et al. 2009). Identical to study 1, we randomized the level of incongruence between the ILE score and the qualitative evaluation.[4]

Based on our theory, we expect that participants will be more inclined to fire the teacher with the lowest ILE score when that teacher has a Moroccan-sounding name ($H_2$). Our sample testing the selective adherence hypotheses were therefore filtered to include only respondents of Dutch descent ($n = 792$).[5] In our analyses below, we examine the effect of this manipulation on participants' inclination to follow the ILE score, and its interaction with the type of advice (algorithmic versus human, $H_3$).

It is important to note that previous vignette survey experimental studies have frequently failed to identify discriminatory patterns, which has been explained by methodological reasons, mainly social desirability pressures and the difficulty of simulating the conditions of real-world decision making (Wulff and Villadsen 2020). We were certainly aware of this limitation when designing our study, and for this reason we argue that our study can be considered as a particularly hard case for our *selective adherence* hypothesis.

Similar to study 1, we excluded from all the analyses participants who did not pass the attention check or completed the questionnaire in less than 3 minutes. These filtering criteria are not associated with the assignment to the experimental conditions (supplementary table A2.2). After this filtering, we were left with an analytical sample of $N = 904$ for the replication of study 1 (automation bias hypothesis) and $N = 792$ for testing of our selective adherence hypotheses, that is, $H_2$ and $H_3$, the teacher ethnicity manipulation. The advice groups and teacher names' groups are balanced in relation to gender, reported income, education, and age (supplementary table A1.2).

## Results (Study 2)

### *Automation Bias*

The results of study 2 with regards to the automation bias hypothesis are displayed in tables 3 and 4. Consistent with our study 1 findings, we find small, statistically insignificant differences between algorithmic-advice and human-advice conditions. Including covariates and restricting the samples to those who successfully completed the manipulation checks does not change the results (supplementary tables A5.1.1 and A5.2.1).

Also, there were no major differences across the randomized incongruence versions. In both cases, the differences are in the expected direction, yet they are relatively small and statistically insignificant (supplementary tables A5.3.2 and A5.3.4).

Thus, in both studies 1 and 2, we did not find that participants are more likely to follow the algorithmic advice compared with equivalent human advice. We also pooled the two samples to maximize statistical power ($N = 1,509$), and the differences, while in the expected direction (11.1% versus 10.5%), were still statistically insignificant ($OR = 1.07$, $z = 0.41$, tables 3 and 4).[6] We do not find support for automation bias. We further ruled out that potential differences in demographic and socioeconomic characteristics between our sample and the civil service population might impact our experimental results via interaction models (supplementary table A5.5.1). We also examined whether participants' propensity to follow the algorithmic advice is influenced by their familiarity with the use of algorithms by public organizations. 21% of the participants assigned to the algorithmic advice in the two studies reported that they were familiar with algorithm use by public bodies. This variable too had an insignificant effect (supplementary table A5.6.1).

### *Selective Adherence*

We now turn to discuss the results of our second study in relation to our hypotheses of *selective adherence*. Table 5 reports the regression results of the comparison between the two teacher ethnicity conditions, across the algorithm and human advice. In Model 1, we regressed our outcome variable on the two manipulations to test their main effects, and thereafter in Model 2 we add their interaction. Table 6 then summarizes the descriptive differences in raw scores.

We find a main effect for the teacher ethnicity manipulation in the expected direction. Respondents are more likely to adhere to an advice when it predicts low performance for a decision subject from a negatively stereotyped minority. A Moroccan-Dutch teacher with a low ILE score is 50% more likely *not* to have their contract renewed, compared to a Dutch teacher with the same score ($OR = 1.50$, $p = .04$, one-sided test). Descriptively, the difference in probabilities is 12.3% versus 8.6%. In other words, in line with our $H_2$, we find *selective adherence across both types of advice*: human and algorithmic. This effect remains positive and significant when controlling for covariates (supplementary table A5.1.2).[7] Given established difficulties for survey experimental designs to identify such discriminatory patterns, these findings are important and likely speak to the prevalence of such biases.

This pattern is consistent across our two incongruence conditions (supplementary tables A5.3.3 and A5.3.5), which further speaks to its robustness. We also examined the interaction between the teacher ethnicity manipulation and participants' age, gender, level of education, and reported income. All these interactions are statistically insignificant (supplementary table A5.5.2).

[4]In study 2, we did not randomize the order of the three teachers, based on the results of study 1.

[5]We assume negative stereotypes toward citizens of migrant descent to be more emphasized among citizens without a migration background.

[6]For this sample size and baseline probability, we estimate that a small effect-size of $OR = 1.45$ (a probability change of approximately 5%) is detectable (power = 0.8, $p = .05$, one-sided test). For post hoc power analyses, see supplementary section A7.

[7]Consistently, the coefficient is positive in supplementary analyses after filtering out those who did not properly read the task ($OR = 1.39$, 9.5% versus 7%), yet it is not sufficiently significant, arguably due to the smaller sample size (supplementary table A5.2.2).

**Table 3.** Study 2—Regression Results of Participants' Adherence to Algorithmic Versus Human-Expert Advice (Automation Bias)

| Predictors | Study 2 (1) | | | Studies 1 and 2 (pooled) (2) | | |
|---|---|---|---|---|---|---|
| | *OR* [95% CI] | *z* | *p* value | *OR* [95% CI] | *z* | *p* value |
| Algorithm | 1.16 [0.75–1.80] | 0.68 | .498 | 1.07 [0.77–1.48] | 0.41 | .683 |
| Study 2 (ref. = study 1) | | | | 0.85 [0.61–1.18] | −0.96 | .335 |
| Intercept | 0.10 [0.08–0.14] | −13.91 | <.001 | 0.13 [0.09–0.17] | −13.57 | <.001 |
| Observations | 904 | | | 1,509 | | |
| Log-likelihood | −297.144 | | | −516.073 | | |

*Note:* Binary outcome: following the ILE score (1 = non-renewal of employment of teacher with lowest ILE score).

**Table 4.** Study 2—Descriptive Results (Automation Bias)

| Outcome: Teacher Selected (Non-renewal of employment) | Study 2 (*N* = 904) | | Studies 1 and 2 (pooled) (*N* = 1,509) | |
|---|---|---|---|---|
| | Algorithmic Advice (*n* = 461), % | Human-Expert Advice (*n* = 443), % | Algorithmic Advice (*n* = 756), % | Human-Expert Advice (*n* = 753), % |
| **Teacher with lowest ILE score (algorithmic/human-expert)** | **10.8** | **9.5** | **11.1** | **10.5** |
| Teacher with poorest qualitative evaluation | 81.6 | 83.3 | 79.9 | 82.3 |
| Other | 7.6 | 7.2 | 9.0 | 7.2 |

Next, we examine the interaction effect, in line with our $H_3$. While we find statistically significant evidence that participants are more inclined to follow the ILE score when the prediction is aligned with stereotypes, our findings do not suggest that this bias is increased when the score is produced by an algorithm compared to human advice, in contrast with our $H_3$. Participants under both conditions were more likely not to renew the contract of the teacher of Moroccan background, and the interaction between the teacher ethnicity manipulation and the algorithmic advice condition is not statistically significant in our interaction model (table 5, model 2). The differences between the Moroccan-Dutch and Dutch teachers in the human-expert advice group were slightly greater compared with the algorithmic group, yet these descriptive differences are not statistically significant, as evidenced by the insignificant interaction term ($|z|$ = 1.45, $p$ = .147),[8] and as such could be entirely due to chance (type I error). The coefficient of the interaction further diminishes when we control for covariates ($|z|$ = 1.25, $p$ = .210), and restricting the samples to those who successfully completed the manipulation checks does not change the results (supplementary tables A5.1.2 and A5.2.2).

On this basis, given the significant main effect and insignificant interaction, our findings indicate that decision-makers are subject to selective adherence when processing decisional aid outputs, regardless of whether these outputs are produced by humans or algorithms. At the same time, and despite our considerable sample size, we acknowledge statistical power limitations in our interaction analysis ($H_3$).[9] Still, we can infer with sufficient confidence that a significant increase (as we hypothesized) is unlikely due to the fact that the interaction coefficient is in the opposite direction.

To summarize our main findings in study 2, we find that participants, across both sources of advice (human and algorithmic), tend to follow the advice *in a selective manner*—when it corresponds to pre-existing biases and stereotypes, which translates into group disparities (in support of our $H_2$). All else constant, a Moroccan-Dutch teacher is significantly more likely to get sanctioned due to a negative evaluation score compared to a Dutch teacher with the same score. Our findings indicate that there are no significant differences between human and algorithmic advice in this respect. Selective biased processing patterns are found for both types of advice, persisting in algorithm adoption.

## Study 3: Replication with a Sample of Civil Servants

In study 3, we aimed to replicate studies 1 and 2 with a sample of civil servants. For this purpose, we contracted a government-owned personnel research program (Internetspiegel/ICTU) operating an online panel of Dutch civil servants (Flitspanel/"Flashpanel"). Participating civil servants register themselves for their participation in the panel, and it is used both by the government itself to survey current policy issues as well as for academic studies.[10]

The online survey was administered and distributed by ICTU to 3,294 civil servants. The fieldwork was conducted between the 2nd and 22nd of February 2021. ICTU sent the invitations to the participants via email, followed by two reminders. A total number of 1,345 participants completed

[8]A similar result is produced by a likelihood ratio comparing the interaction model with a main-effect-only model ($\chi^2(1)$ = 2.141, $p$ = .143). Comparing the two models via BIC and AIC indicates that the main effect model is more appropriate (Lorah 2020).

[9]See supplementary section A7.

[10]For additional information, see https://flitspanel.nl.

[11]Assuming the baseline probability we found in studies 1 and 2, we estimate this sample size is sufficient for detecting a modest effect-size of $OR$ = 1.52.

**Table 5.** Regression Results of Study 2—Participants' Selective Adherence to Advice (Algorithmic Versus Human-Expert) that Matches Stereotypical View of Decision Subjects

| Predictors | (1) | | | (2) | | |
|---|---|---|---|---|---|---|
| | *OR* [95% CI] | *z* | *p* value | *OR* [95% CI] | *z* | *p* value |
| Algorithm | 1.20 [0.76–1.91] | 0.78 | .438 | 1.80 [0.88–3.83] | 1.58 | .114 |
| Moroccan-Dutch teacher | 1.50 [0.95–2.40] | 1.73 | .083 (.042 *one-sided*) | 2.23 [1.11–4.73] | 2.19 | 0.029 |
| Algorithm × Moroccan-Dutch teacher | | | | 0.50 [0.19–1.27] | −1.45 | .147 |
| Intercept | 0.08 [0.05–0.13] | −11.16 | <.001 | 0.07 [0.03–0.11] | −9.10 | <.001 |
| Observations | 792 | | | 792 | | |
| Log-likelihood | −261.789 | | | −260.719 | | |
| BIC | 543.602 | | | 548.136 | | |
| AIC | 529.579 | | | 529.438 | | |

*Note:* Binary outcome: following the ILE score (1 = non-renewal of employment of teacher with lowest ILE score).

**Table 6.** Study 2—Descriptive Results (Selective Adherence)

| Outcome: Teacher Selected (Non-renewal of employment) | All (*n* = 792) | | Algorithmic Advice (*n* = 405) | | Human-Expert Advice (*n* = 387) | |
|---|---|---|---|---|---|---|
| | Teacher with lowest ILE score: | | Teacher with lowest ILE score: | | Teacher with lowest ILE score: | |
| | Dutch (*n* = 409), % | Moroccan-Dutch (*n* = 383), % | Dutch (*n* = 216), % | Moroccan-Dutch (*n* = 189), % | Dutch (*n* = 193), % | Moroccan-Dutch (*n* = 194), % |
| **Teacher with lowest ILE score (algorithmic/human-expert)** | **8.6** | **12.3** | **10.6** | **11.6** | **6.2** | **12.9** |
| Teacher with poorest qualitative evaluation | 84.1 | 79.9 | 82.4 | 78.8 | 86.0 | 80.9 |
| Other | 7.3 | 7.8 | 6.9 | 9.5 | 7.8 | 6.2 |

the survey (41% response rate).[11] The sample includes civil servants working in different policy sectors, at both national and local levels.[12] Yet, it should be noted that the sample is not entirely representative of the Dutch public sector. Women are underrepresented in our sample, and the mean age was higher compared with the Dutch public sector. In Appendix 2, we present the demographic characteristics of the sample and control for these characteristics in our robust analyses, as detailed below.

We repeated the 2 × 2 factorial design and the experimental procedure of study 2. The online survey was administered by ICTU (using a different software than Qualtrics), and for technical reasons we could not include the additional randomization into high and modest incongruence. Hence, in this study, we assigned all participants to the high incongruence scenario, where the teacher who receives the lowest ILE score is the one with the most positive qualitative evaluation. We did not include an attention check in this survey, as per the panel's request, and therefore our analytical sample for the automation bias hypothesis is the full sample of 1,345 participants.[13] For the analyses of the teacher's ethnicity manipulation, same as in study 2, we included only participants of Dutch descent (*N* = 1,203). This screening is not associated with the assignment to the experimental conditions (supplementary table A2.3), and the randomization groups are balanced in relation to gender, reported income, education, and age (supplementary table A1.3).

It is important to note that the fieldwork of study 3 coincided with the occurrence of significant events in the Netherlands, surrounding the "childcare benefits scandal" (*toeslagenaffaire* in Dutch). The scandal reached its peak during the technical preparations of our survey, and shortly before its distribution with growing public attention in December 2020, following the publication of a parliamentary report on the scandal (titled "Unprecedented Injustice," Parlementaire Ondervragingscommissie Kinderopvangtoeslag 2020), resulting in the resignation of the Dutch government mid-January 2021.

The scandal involved the reliance by the Dutch tax authorities on an AI algorithm—a "learning algorithm" that used, among other criteria, nationality as a discriminant predictive feature, and served as a decisional aid in flagging high-risk applicants for further scrutiny. The requests flagged by the algorithm were checked manually by tax employees after considering (and/or requesting from applicants) additional information (Autoriteit Persoonsgegevens/Dutch Data Protection Authority 2020). The scandal disproportionately affected citizens of foreign descent, with mostly dual nationality families wrongly accused of benefits fraud: "[T]he tax ministry singled out tens of thousands of families often on the basis of their ethnic background" (Financial Times 2021). Victims of the scandal were required to retroactively repay large sums of money (amounting to as much as tens of thousands of euros), with the financial strain reportedly resulting in acute financial problems, bankruptcies, mental health issues, and broken families (Geiger 2021).

[12]The survey was sent to participants from the following sectors: central government (national ministries), municipalities, provinces, inter-municipal cooperative arrangements, water boards, defense, and police.

[13]All participants completed the survey in more than 3 minutes.

The scandal is a textbook example of the meeting point between algorithmic bias and human decision-makers' biases. While the system itself was biased, using nationality as a predictive feature, the way tax officials went about their work reinforced the system's biases: "Both the automated risk selection and the individual investigations of officials were discriminatory, the data protection authority ruled" (Volkskrant 2020). The scandal is illustrative of the patterns diagnosed in our study 2: algorithmic recommendations aligned with prevalent stereotypes (i.e., indicating a negative assessment for members of an ethnic minority group) with decision-makers likely not to override such recommendations. Victims of the scandal, much like our teacher of Moroccan heritage in study 2, were specifically singled out for targeted scrutiny because of their ethnic origin or double nationality, following an algorithmic prediction ("families of largely Moroccan, Turkish and Dutch Antilles origin were targeted, according to the national data protection authority," Financial Times 2021).

Given that the survey was conducted shortly after the scandal, the results of this study should be interpreted in light of it. Our participants were highly aware of the risk of algorithmic bias, and sensitive to this issue. 63% of the participants reported that they are familiar with the use of algorithms by public organizations, and more than half of these (33%) mentioned this case when asked to give an example, and many of them spontaneously expressed their criticism toward it in their qualitative answers. While we anticipated such public reactions to be reflected in participants' answers, we decided not to withhold the fieldwork, as we believe that investigating our research question under these conditions can yield meaningful insights. We return to this point below in our discussion.

## Results—Study 3

### *Automation Bias*

The logistic regression results of study 3 with regards to the comparison between algorithmic and human advice (automation bias hypothesis) are displayed in table 7, with descriptive differences presented in table 8. Participants were significantly less likely to follow the ILE score when produced by an algorithm, and more likely to select the teacher with the poorest qualitative evaluation. This confirms the findings of our previous two studies, which similarly did not diagnose automation bias in decision making. In fact, the patterns in study 3 are in the opposite direction to the automation bias expectations. Including covariates and restricting the sample to those who successfully completed the manipulation checks does not change the results (supplementary tables A5.1.1 and A5.2.1).

Also, the interactions between the advice manipulation and participants' gender, age and higher education are all statistically insignificant (supplementary table A5.5.3). This suggests that a sample more representative of the civil service's demographic and socioeconomic characteristics would have yielded similar results. However, in contrast with studies 1 and 2, the negative effect of the algorithmic advice is linked to respondents' reported familiarity with the use of algorithms by public organizations. When filtering our sample in study 3 to participants who were not familiar with the use of algorithms by public organizations before the survey ($n$ = 498), the likelihood of following the algorithmic advice increases and is not significantly lower compared to the human-expert condition (6.9% versus 8.8%, $p$ = .431, two-sided). We return to this point later in our discussion.

**Table 7.** Study 3—Regression Results of Participants' Adherence to Algorithmic Versus Human-Expert Advice (Automation Bias)

| Predictors | (1) | | |
|---|---|---|---|
| | *OR* [95% CI] | *z* | *p* value |
| Algorithm | 0.54 [0.34–0.83] | −2.74 | .006 |
| Intercept | 0.09 [0.07–0.12] | −17.32 | <.001 |
| Observations | 1345 | | |
| Log-likelihood | −329.022 | | |

*Note:* Binary outcome: following the ILE score (1 = non-renewal of employment of teacher with lowest ILE score).

**Table 8.** Study 3—Descriptive Results (Automation Bias)

| Outcome: Teacher Selected (Non-renewal of employment) | Algorithmic Advice ($n$ = 662), % | Human Advice ($n$ = 683), % |
|---|---|---|
| **Teacher with lowest ILE score (algorithmic/human-expert)** | 4.8 | 8.6 |
| Teacher with poorest qualitative evaluation | 89.1 | 83.9 |
| Other | 6.0 | 7.5 |

### *Selective Adherence*

Tables 9 and 10 present the results of the comparison between the teacher ethnicity conditions, across the algorithm and human advice, which is relevant to our selective adherence hypotheses. We find a negative main effect for the Moroccan-Dutch teacher (table 9). In departure from our study 2, participants in study 3 (civil servants in the aftermath of a major public scandal involving algorithm use and ethnic discrimination) were less likely to fire a Moroccan-Dutch teacher with a low ILE score, compared to a Dutch teacher with the same score. These differences are fairly similar across the two groups, and the interaction is insignificant. These results do not change when adding controls and filtering out those who did not properly read the task (supplementary tables A5.1.3 and A5.2.3).

To summarize the main results, in this study with a sample of civil servants, similar to our previous two studies, we do not find support for automation bias. Our study 3 reveals participants in the aftermath of the scandal were less likely to be influenced in their decision by the ILE score when generated by an AI algorithm rather than by human experts. Also, and in contrast with our study 2, they were less likely to sanction the Moroccan-Dutch teacher, regardless of the type of advice. These findings arguably speak to the effect of the scandal in shaping bureaucratic responses, as we discuss below.

**Table 9.** Regression Results of Study 3—Participants' Selective Adherence to Advice (Algorithmic Versus Human-Expert) that Matches Stereotypical View of Decision Subjects

| Predictors | (1) | | | (2) | | |
|---|---|---|---|---|---|---|
| | *OR* [95% CI] | *z* | *p* value | *OR* [95% CI] | *z* | *p* value |
| Algorithm | 0.46 [0.28–0.74] | −3.12 | .002 | 0.53 [0.28–0.95] | −2.08 | .038 |
| Moroccan-Dutch teacher | 0.57 [0.35–0.91] | −2.33 | .020 | 0.64 [0.36–1.13] | −1.52 | .129 |
| Algorithm × Moroccan-Dutch teacher | | | | 0.69 [0.24–1.89] | −0.71 | .480 |
| Intercept | 0.13 [0.09–0.17] | −12.13 | <.001 | 0.12 [0.08–0.17] | −11.36 | <.001 |
| Observations | 1,203 | | | 1,203 | | |
| Log-likelihood | −286.297 | | | −286.044 | | |
| BIC | 593.872 | | | 600.458 | | |
| AIC | 578.594 | | | 580.087 | | |

*Note:* Binary outcome: following the ILE score (1 = non-renewal of employment of teacher with lowest ILE score).

**Table 10.** Study 3—Descriptive Results (Selective Adherence)

| **Outcome: Teacher Selected (Non-renewal of employment)** | **All (*n* = 1,203)** | | **Algorithmic Advice (*n* = 595)** | | **Human-Expert Advice (*n* = 608)** | |
|---|---|---|---|---|---|---|
| | **Teacher with lowest ILE score:** | | **Teacher with lowest ILE score:** | | **Teacher with lowest ILE score:** | |
| | **Dutch (*n* = 603), %** | **Moroccan-Dutch (*n* = 600), %** | **Dutch (*n* = 303), %** | **Moroccan-Dutch (*n* = 292), %** | **Dutch (*n* = 300), %** | **Moroccan-Dutch (*n* = 308), %** |
| **Teacher with lowest ILE score (algorithmic/human-expert)** | **8.3** | **5.0** | **5.9** | **2.7** | **10.7** | **7.1** |
| Teacher with poorest qualitative evaluation | 83.6 | 89.2 | 86.8 | 92.5 | 80.3 | 86.0 |
| Other | 8.1 | 5.8 | 7.3 | 4.8 | 9.0 | 6.8 |

## Discussion and Conclusion

With AI set to fundamentally alter decision making in public organizations, how do human decision-makers actually process algorithmic advice? Drawing on two separate strands of behavioral literature, we have theorized that two biases in particular are of high relevance and in dire need of investigation by public administration scholars: "automation bias" and "selective adherence" to algorithmic advice.

A first bias stemming from automation studies is that decision-makers would automatically default to the algorithm, potentially then also to poor algorithmic advice, ignoring contradictory cues from other sources: *automation bias*. A second hypothesized bias, which we extrapolated from public administration literature, regards decision-makers' tendency to defer to the algorithm selectively—when algorithmic predictions match pre-existing stereotypes: *selective adherence*. The use of algorithms could then disproportionately negatively affect stereotyped groups, potentially creating administrative burdens (Herd and Moynihan 2019) and compounding discrimination. Below we discuss and reflect, in turn, on our findings in relation to each of these biases and their implications for public sector decision making in the age of automation.

### Automation Bias

Overall, our experimental findings from three separate studies with an aggregated sample of 2,854 participants do not reveal a general pattern of automatic adherence to algorithmic advice. Across the three studies, we consistently did not find evidence for an overall tendency for *automation bias*.

In none of the three studies were participants more likely to follow the ILE score when produced by an algorithm compared to a human-expert: in studies 1 and 2, the differences were small and statistically insignificant, and in study 3 conducted shortly after the childcare benefits scandal, participants were actually less likely to follow the algorithmic advice, indicative of a growing reluctance to trust algorithms in its aftermath. We attribute this latter negative effect primarily to the proximity of the study to the scandal, increasing participants' exposure to the dangers of reliance on AI algorithmic models (as exemplified by the scandal). A considerable number of respondents in study 3 (33%) were aware of the use of algorithms in the benefits scandal, as evidenced by their open answers. Furthermore, as reported above, among those respondents who were not aware of the use of algorithms by public organizations we did not find a lower propensity to follow the algorithmic advice compared to human advice. This suggests the results of study 3 represent a response to the scandal rather than indicative of an inherent distrust toward algorithmic-sourced advice. At the same time, our study should also serve as further caution as to the adoption of unvetted, under-performing algorithmic systems in public sector decision making, increasingly diagnosed in practice (e.g., Ferguson 2017; Eubanks 2018; O'Neil 2016), and as exemplified in our article by the Dutch childcare benefits scandal. Such failures, once exposed, have consequences, with poorly implemented systems resulting in lower levels of trust in algorithms' performance.

These experimental findings are largely consistent with findings from earlier studies outside our discipline on pretrial algorithmic risk scores in the US context. These studies, too, did not reveal an overwhelming pattern of automatic adherence to algorithmic risk scores. An important limitation of these previous studies however, was that they failed to compare algorithmic advice with equivalent human advice, which we remedy with our current investigation.

Still, how can we reconcile the results of our study (and the studies above) with findings from studies in social psychology on the use of automation in aviation and healthcare (e.g., Lyell and Coiera 2017; Skitka, Mosier, and Burdick 1999, 2000), where patterns of automation bias have been well-documented and recognized? One possible explanation for this discrepancy is a relative skepticism about the performative capacity of AI algorithms, with many participants, based on their self-reporting, still under-exposed to their performative capacities (in studies 1 and 2), or exposed to their negative consequences (in study 3, following the benefits scandal). This is an important difference to earlier studies on automation applied in areas well-accustomed to such devices (aviation, healthcare), characterized by routine use of reliable automation, resulting in high levels of trust in their performance.

These findings also have important implications for the public administration literature on automation and discretion. Introducing algorithmic tools into the decision-making process, we find in our studies, did not supplant the discretion and judgment of human decision-makers, with the vast majority of our respondents overriding the prediction. At the same time, we argue that it is too soon to rule out concerns with undue bureaucratic deference to AI systems. Rather, automatic deference to algorithmic advice could become more prevalent as decision-makers become increasingly exposed to AI algorithms in the practice of public organizations. Repeated experience with high-performing systems (in so far as such systems are high-performing) might increase "user appreciation" of their judgment capacities (decrease skepticism), leading to higher levels of deference over repeated interactions.

## Selective Adherence

We also theorized, extrapolating from behavioral literature, that similar to human advice, decision-makers are likely to rely on algorithmic inputs in a biased, selective manner—to assign more weight to the advice and follow it against contradicting evidence when this is aligned with pre-existing stereotypes. Establishing whether selective adherence is present across both types of advice is important in a context where AI algorithms are said to have the potential to do away with human decisional biases. We further theorized that selective adherence biases could be exacerbated by algorithms, by virtue of their unique nature.

In study 2, which consisted of a sample of Dutch citizens in a context where citizens can serve as actual decision-makers, we found evidence supporting *selective adherence* patterns across both human and algorithmic advice conditions. Namely, when the low prediction score is assigned to a teacher from a negatively stereotyped ethnic minority, participants were significantly more likely to rely on it in their decisions and less likely to override it. These selective adherence patterns are present across both types of advice (human and algorithmic), as evidenced by the positive and significant main effect for the teacher ethnicity manipulation. In both conditions, participants were more likely not to renew the contract of the ethnic minority teacher.

Importantly, we found that this bias is not more emphasized for algorithmic advice when compared to human advice, as the interaction between the two manipulations in our factorial design is insignificant. Taken jointly, these two sets of findings indicate that while not exacerbated by algorithms, selective adherence patterns occur across both sources of advice rather than being restricted to human advice. The replacement of human advice with algorithmic advice does not make selective adherence disappear. These findings are also in line with results from other studies on pretrial algorithmic risk scores by law and computer science scholars respectively, which report patterns consistent with biased adherence to algorithmic advice (Green and Chen 2019a, 2019b; Stevenson 2018). The findings of our study, and others, carry important implications as they indicate *decision-making biases endure in AI algorithm adoption* as decisional aides in the public sector, contrary to the promise that propelled their adoption as a means to do away with such biases. Similar to human-sourced advice, a tendency to follow algorithmic advice, too, rather than generalized, is instead *selective* and more likely when this advice matches pre-existing stereotypical beliefs. Our sample of civil servants in study 3 did not yield similar results. The participants in study 3, conducted in the aftermath of a scandal involving algorithm use and discrimination in bureaucratic decision making, were less likely not to renew the contract of a teacher from a negatively stereotyped minority with a low score compared to a Dutch teacher with the same score.

How can we explain the discrepancy between the studies on this aspect? Several explanations could account for these findings: First, one could speculate that these differences might stem from distinctive characteristics of civil servants compared with lay citizens, namely the ability of the former to overcome social biases and prejudice as a result of their professional training, expertise, or background. However, a vast body of literature in social science provides us with theory and empirical evidence for the existence of discriminatory decision-making that are also rooted in subtle and unconscious cognitive mechanisms (e.g., Schram et al. 2009). These patterns have been theorized and are well-documented in bureaucratic contexts, also among highly educated, professional decision-makers (e.g., Andersen and Guul 2019; Assouline, Gilad, and Bloom 2022; Giulietti, Tonin, and Vlassopoulos 2019).

A methodological explanation is also plausible for the patterns encountered in study 3: namely, that civil servant participants' responses could reflect social desirability bias, an (unconscious) need to answer questions in ways that demonstrate that they do not discriminate. This is a common threat to studies of discrimination more broadly, and indeed several survey experimental studies have "failed" to find racial discrimination in their data, arguably for this reason (e.g., Baekgaard and George 2018; Wulff and Villadsen 2020). This threat is plausibly more likely for the sample of professional civil servants surveyed in study 3, compared with the sample of lay citizens in study 2 (even though both groups were explicitly guaranteed anonymity). Furthermore, this threat is potentially exacerbated by the fact that civil servant participants were invited by a panel linked to the Dutch government.

The more plausible explanation, in our reading, for the fact that we did not encounter patterns of selective adherence in study 3 (as we did in study 2) is that participants' responses were an authentic reaction to the recent childcare benefits scandal and the political, media and public scrutiny that followed from it. The scandal represented a case of systemic bureaucratic discrimination against citizens with a migration background, an empirical case that incidentally closely matched our own hypothetical scenario, with many civil servants respondents spontaneously indicating familiarity with the scandal in their open answers. It is likely that the scandal increased civil servants' awareness of racial profiling and discrimination toward ethnic minorities in the Netherlands—explicitly also in relation to algorithm use in bureaucratic decision-making. Indeed, social psychology studies have theorized that racial biases can be attenuated when people are highly motivated to do so (Devine et al. 2002). This would suggest the scandal had a learning effect although our study does not allow us to assess to what extent these effects are long-lived.

It is important to note that the scandal itself is an illustrative example of the theorized patterns of decision-makers' adherence to algorithmic advice, and how it can result in discrimination in decision making. The scandal speaks acutely to the serious real-life repercussions that can arise when human bias meets algorithmic bias in bureaucratic decision making. Taken together with our empirical findings in study 2, we believe that there is evidence for selective adherence to algorithmic advice that calls for additional and pressing investigation of this issue by public administration scholars.

A key justification put forward for algorithm adoption in high-stakes public sector areas such as criminal justice or policing, and for "tolerating" shortcomings of such systems (e.g., pertaining to their opaqueness and associated concerns with transparency and accountability) has been their perceived superior performance and said "objectivity" as data-driven technologies, as a way to overcome human biases and limitations. While such claims have been deflated when it comes to algorithms' own learning and functioning (e.g., algorithms replicating and propagating systematic biases learned from training data is a well-documented problem that can arise in algorithm deployment), it is important to keep in mind that bias can also crop up at another level: in the human–AI interaction, in how decision-makers process, interpret, and act upon algorithmic outputs. Our findings raise further questions about the added value of the reliance on algorithmic advice as a mechanism to avoid bias and speak to potential negative effects of automation of the administrative state for already vulnerable and disadvantaged citizens (Eubanks 2018; Ranchordas 2022). Even assuming that the algorithmic outputs themselves could be bias-free, we find some evidence that human decision-makers tend to rely on such outputs selectively, that is, when their predictions "suit" pre-existing stereotypes.

Keeping humans-in-the-loop (human intervention) is an important safeguard against algorithmic failures and is even legally mandated to that end in forward-looking regulatory frameworks such as the EU GDPR. While our findings as to a lack of automatic deference are encouraging in this context, the likelihood that decision-makers adhere to algorithmic advice (rather than resist it) *precisely* when predictions are aligned with group stereotypes and disadvantage minority groups is disconcerting, and speaks to potential blind spots in our ability to exercise meaningful oversight. Such concerns can become especially problematic, as we saw, in mixed algorithmic decision making when human bias meets algorithmic bias. At the same time, an encouraging, tentative take-away that emerges from our investigation is that high-visibility, public exposure of such biases (as in the aftermath of the benefits scandal) can have learning effects through rendering civil servants more conscious and alert to such risks leading to their potential attenuation in decision making, at least in the short term.

Our study takes a first step to investigate how public decision-makers process AI algorithmic advice from decisional support systems. As AI tools proliferate in the public sector, this comes with significant possible implications for the nature of administrative decision making, rendering this issue increasingly salient for our discipline. Future studies may investigate these aspects in scenarios pertaining to different sectors and across multiple national jurisdictions. Importantly, and following our results, follow-up work could further test the role of decision-makers' learning and repeat exposure through a design that allows for repeat interactions with the algorithm so as to assess to what extent participants' trust in the algorithm changes over time, potentially leading to patterns of enhanced deference. Investigating the cognitive mechanisms underpinning algorithmic decision making in an administrative context will be of crucial theoretical and empirical significance, part and parcel of tackling broader, fundamental questions as to the impact of artificial intelligence for bureaucratic expertise and discretion, the nature of public authority, and public accountability in the age of automation.

## Supplementary Material

Supplementary data is available at the *Journal of Public Administration Research and Theory* online.

## Funding

This article is part of a project that has received funding from the European Research Council (ERC) under the European Union's Horizon 2020 research and innovation programme (grant agreement 716439).

## Data Availability

The data underlying this article are available in Harvard Dataverse, at https://doi.org/10.7910/DVN/TQYJNF.

## Appendix A

### Randomization Groups

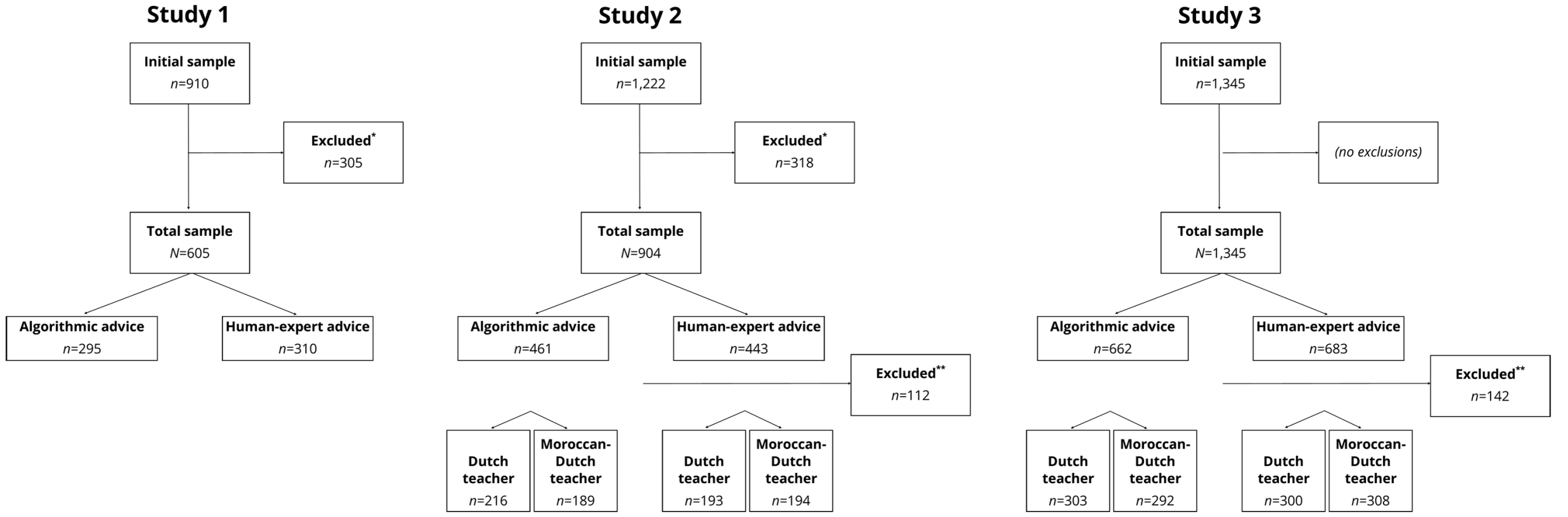

Exclusions:

* Participants who failed the attention check or completed the questionnaire in less than 3 min.
** Participants not of Dutch descent (excluded from the analysis of selective adherence hypotheses—$H_2$ and $H_3$).

## Appendix B

### Sample Characteristics

| | | Study 1 | Study 2 | Study 3 | Dutch Civil Service |
|---|---|---|---|---|---|
| Gender | % Women | 54.5 | 47.8 | 28.5 | 42.0 |
| Age | Mean (SD) | 47.1 (16.6) | 47.5 (17.5) | 55.4 (7.5) | 46.3 |
| | % | | | | |
| | 18–25 | 14.3 | 17.3 | 0.1 | 19.7 |
| | 26–35 | 14.6 | 12.6 | 1.4 | (age 18–35) |
| | 36–45 | 17.3 | 12.8 | 9.1 | 50.0 |
| | 46–55 | 18.0 | 18.5 | 33.4 | (age 36–55) |
| | 56–65 | 19.7 | 21.4 | 54.3 | 30.3 |
| | 65+ | 16.0 | 17.4 | 1.6 | (age 56+) |
| Education | % high education | 50.1 | 49.0 | 71.0 | 50.0 |
| *N* participants | | 605 | 904 | 1,345 | |

*Note*: Valid percentages are reported. Dutch civil service data is from 2018, regarding 412,999 civil servants from national and local civil service, including defense and police.
*Source:* https://kennisopenbaarbestuur.nl/.